\documentstyle[12pt,epsf,]{article}

\newcommand{\beq}{\begin{equation}}
\newcommand{\eeq}{\end{equation}}
\newcommand{\beqa}{\begin{eqnarray}}
\newcommand{\eeqa}{\end{eqnarray}}
\newcommand{\no}{\nonumber}
\newcommand{\q}{\quad}
\newcommand{\qq}{\qquad}
\newcommand{\mnod}{\stackrel{\circ}{M}}
\newcommand{\Fnod}{\stackrel{\circ}{F}}
\newcommand{\tr}{\mbox{tr}}

\begin{document}

\hfill 

\hfill 

\bigskip\bigskip

\begin{center}

{{\Large\bf The role of resonances in non--leptonic hyperon decays}}

\end{center}

\vspace{.4in}

\begin{center}
{\large Bu\={g}ra Borasoy\footnote{email: borasoy@het.phast.umass.edu}
 and Barry R. Holstein\footnote{email: holstein@phast.umass.edu }}

\bigskip

\bigskip

Department of Physics and Astronomy\\
University of Massachusetts\\
Amherst, MA 01003, USA\\

\vspace{.2in}

\end{center}

\vspace{.7in}

\thispagestyle{empty} 

\begin{abstract}
We examine the importance of resonances for the non--leptonic hyperon
decays in the framework of chiral perturbation theory. Lower lying
resonances are included into the effective theory. Integrating
out the heavy degrees of freedom in the resonance saturation scheme 
generates higher order counterterms in the effective Lagrangian,
providing an estimate of the pertinent coupling constants.
A fit to the eight independent decay amplitudes that are not related
by isospin symmetry is performed and reasonable agreement for {\it both}
s-- and p-- waves is achieved. 
\end{abstract}

\vfill

\section{Introduction}
The matrix elements of non--leptonic hyperon decays can be described in terms
of just
two amplitudes --- the parity--violating s--wave and the parity--conserving
p--wave. 
Chiral perturbation theory provides a framework
whereby these amplitudes can be expanded in terms of small
four--momenta and the current masses $m_q$ of the light quarks, $q=u,d,s$.
At lowest order in this expansion 
the amplitudes are expressed in terms of two unknown coupling
constants, so--called low energy constants (LECs). However, there is no 
consensus for the determination of these two weak parameters. 
If one employs values which provide a good fit for the s--waves, one obtains
a poor description of the p--waves. On the other hand, a good p--wave 
representation yields a poor s--wave fit \cite{DGH}.
In order to overcome this problem, one must go beyond leading order.
In the paper of Bijnens et al. \cite{BSW}, a first attempt was
made in calculating the leading chiral corrections to such decays.
The authors worked in the limit $m_u=m_d=0$ and kept only the non--analytic
logarithms from the Goldstone boson loops --- no local counterterms
were considered. However, the resulting s--wave predictions no longer
agreed with the data, and the corrections for the p--waves were even larger.

Jenkins reinvestigated this topic within the framework of heavy baryon
chiral perturbation theory, explicitly including  the spin--3/2 decuplet
in the effective theory \cite{J}. As in \cite{BSW}, no counterterms
were included --- only leading non--analytic pieces from the meson loops
were retained and  $m_u = m_d = 0$ was assumed.
Working in the $SU(6)$ limit (by neglecting the octet--decuplet mass 
splitting), she found significant cancellations between the octet and
decuplet components in the loops. For the s--waves good agreement between
theory and experiment was restored, although in the case of the p--waves,
the chiral corrections did not provide a good description of the data.
Thus, the inability to fit s-- and p--waves simultaneously remains
even after including the lowest non--analytic contributions.

In our recent paper \cite{BH} a calculation was performed which included
{\it all} terms at one--loop order. This work suffers from the fact, however,
that at this order too many new unknown LECs enter the calculation so that
the theory lacks predictive power. In order to proceed these parameters 
were estimated  by means of spin--3/2 decuplet resonance exchange. The results
for the p--waves were still in disagreement with the data and,
therefore, additional counterterms that were not saturated by the
decuplet had to be taken into account. An exact fit to the data was
possible, but the question remains whether the LECs are estimated correctly.

Another intriguing possibility was examined by Le Yaouanc et al., who assert 
that a reasonable fit for both s-- and p--waves can be provided
by appending pole contributions from $SU(6)$ (70,$1^-$) states
to the s--waves \cite{LeY}. Their calculations were performed in a simple
constituent quark model and appear to be able to provide a resolution of the 
s-- and p--wave dilemma.

The purpose of the present work is to consider the validity of this approach
within the framework of chiral perturbation theory. This would provide
also an estimate of the counterterms involved in such a calculation, which
have been neglected completely in \cite{BSW} and \cite{J}.
Furthermore, we include the octet of spin--parity 1/2$^+$ Roper--like states,
which generalizes the considerations of \cite{LeY}.
We do not intend to provide a definitive solution of the problem
of hyperon decay but rather to study the relevance of 
resonance saturation estimation of
counterterms.
We will show that one is able to successfully identify 
counterterms in chiral perturbation theory
with the contributions found in the quark model.
The calculations are performed at tree level. 
Of course, for a more quantitative statement one has 
to include loop effects. However, this is beyond the scope of the present work.

In the following section then, we introduce the 
effective weak and strong Lagrangians including resonant
states. By integrating out the heavy degrees of freedom from the 
theory the effects of the resonances are included in the counterterms and
expressions for the decay amplitudes in terms of these constants are given.
A least--squares fit to experiment is performed in Sec.~3,
while in Sec.~4 we conclude with a brief summary.

\section{Resonances in hyperon decays}
There exist seven experimentally accessible non--leptonic hyperon decays: 
$\Sigma^+  \rightarrow n \, \pi^+ \, , \, 
\Sigma^+  \rightarrow p \, \pi^0 \, , \, 
\Sigma^- \rightarrow n \, \pi^- \, , \, 
\Lambda \rightarrow p \, \pi^- \, ,  \,
\Lambda \rightarrow n \, \pi^0 \, , \,
\Xi^- \rightarrow  \Lambda \, \pi^-\,\mbox{and} \, 
\Xi^0 \rightarrow  \Lambda \, \pi^0 $,
and the matrix elements of these decays can each be expressed in terms of a
parity--violating s--wave amplitude $ A_{ij}^{(S)}$
and a parity--conserving p--wave amplitude $ A_{ij}^{(P)}$
\beq
{\cal A}( B_i \rightarrow B_j \, \pi) =
\bar{u}_{B_j} \Big\{ \, A_{ij}^{(S)} + \, A_{ij}^{(P)}\gamma_5 \Big\}u_{B_i}
\qq .
\eeq
The underlying strangeness--changing Hamiltonian
transforms under $SU(3) \times SU(3)$ as $(8_L, 1_R) \oplus (27_L,1_R)$
and, experimentally, the octet piece dominates over the 27--plet by a factor
of twenty or so. Consequently, we will neglect the 27--plet in what follows.
Isospin symmetry of the strong interactions implies then the relations 
\beqa \label{iso}
&&
{\cal A}(\Lambda \rightarrow p \, \pi^-)
+ \sqrt{2} \, {\cal A}(\Lambda \rightarrow n \, \pi^0) = 0 \no \\
&&
{\cal A}(\Xi^- \rightarrow  \Lambda \, \pi^-)
+ \sqrt{2} \, {\cal A}(\Xi^0 \rightarrow  \Lambda \, \pi^0) = 0 \no \\
&&
\sqrt{2} \, {\cal A}(\Sigma^+ \rightarrow p \, \pi^0)
+ {\cal A}(\Sigma^- \rightarrow n \, \pi^-)
- {\cal A}(\Sigma^+ \rightarrow n \, \pi^+) = 0  \qq ,
\eeqa
which hold for both s- and p-waves.
We choose
$\Sigma^+  \rightarrow n \, \pi^+ \, , \, 
\Sigma^- \rightarrow n \, \pi^- \, , \, 
\Lambda \rightarrow p \, \pi^- \, \mbox{and} \,
\Xi^- \rightarrow  \Lambda \, \pi^- $
as the four independent decay amplitudes which are not related by isospin.

The purpose of this work is to study the role of resonances in 
hyperon decays. To this end, it is sufficient 
in this preliminary study to work at tree level.  
We consider first the Lagrangian
without resonances, which will be included in the following section.
Our starting point is the relativistic effective chiral 
strong interaction Lagrangian for the 
pseudoscalar bosons coupled to the lowest--lying 1/2$^+$ baryon octet
\beq
{\cal L}_{\phi B}^{(1)}  
 = 
 i \, \tr \Big( \bar{B} \gamma_{\mu} [  D^{\mu} , B] \Big) -
\mnod \, \tr \Big( \bar{B} B \Big)
+\frac{1}{2} D \, \tr \Big( \bar{B} \gamma_{\mu} \gamma_5\{ u^{\mu}, B\} \Big) 
+\frac{1}{2}F \, \tr \Big( \bar{B}\gamma_{\mu} \gamma_5 [ u^{\mu}, B]\Big)\q,
\eeq
where the superscript denotes the chiral order and $\mnod$ is the octet
baryon mass in the chiral limit. We set
$ D = 3/4 $ and $ F = 1/2 $,
which are the $SU(6)$  values.
The pseudoscalar Goldstone fields ($\phi = \pi, K, \eta$) are collected in
the  $3 \times 3$ unimodular, unitary matrix $U(x)$, 
\begin{equation}
 U(\phi) = u^2 (\phi) = \exp \lbrace 2 i \phi / \Fnod \rbrace  \qq ,\qq
u_{\mu} = i u^\dagger \nabla_{\mu} U u^\dagger
\end{equation}
with $\Fnod$ being the pseudoscalar decay constant in the chiral limit, and
\begin{eqnarray}
 \phi =  \frac{1}{\sqrt{2}}  \left(
\matrix { {1\over \sqrt 2} \pi^0 + {1 \over \sqrt 6} \eta
&\pi^+ &K^+ \nonumber \\
\pi^-
        & -{1\over \sqrt 2} \pi^0 + {1 \over \sqrt 6} \eta & K^0
        \nonumber \\
K^-
        &  \bar{K^0}&- {2 \over \sqrt 6} \eta  \nonumber \\} 
\!\!\!\!\!\!\!\!\!\!\!\!\!\!\! \right) \, \, \, \, \,  
\end{eqnarray}
represents the contraction of the pseudoscalar fields with the Gell-Mann
matrices.
We replace $\Fnod$ by the pion decay constant $F_\pi \simeq 92.4$ MeV
which is consistent to the order we are working.
$B$ is the standard $SU(3)$ matrix representation of the low--lying
spin--1/2 baryons $( N, \Lambda, \Sigma, \Xi)$ and we work in the 
isospin limit $m_u = m_d = \hat{m}$.

The purely mesonic component of the Lagrangian can be decomposed into a strong
and a weak interacting part
\beq
{\cal L}_{\phi}  = {\cal L}_{\phi}^{S}  + {\cal L}_{\phi}^{W} 
\eeq
where $ {\cal L}_{\phi}^{S} = {\cal L}_{\phi}^{S \: (2)}$ 
is the usual (strong and electromagnetic) mesonic Lagrangian
at lowest order --- cf., {\it e.g.}, \cite{GL}.
From the weak mesonic Lagrangian only the term
\beq \label{weak}
{\cal L}_{\phi}^W = \frac{F_\pi^2}{4} \, h_{\pi} \, {\rm tr}
      \Big( h_+ u_{\mu} u^{\mu}\Big)
\eeq
contributes to the order we are working. 
Here, we have defined
\beq
h_+ = u^{\dagger} h u +  u^{\dagger} h^{\dagger} u  \qquad , 
\eeq
with
$h^{a}_{b} = \delta^{a}_{2} \delta^{3}_{b}$ being
the weak transition matrix. 
Note that $h_+$   transforms as a matter field and 
the weak coupling $h_{\pi}$ is well-determined from
weak non--leptonic kaon decays --- $h_{\pi} = 3.2 \times 10^{-7}$.

Turning to the weak component of the meson--baryon Lagrangian,
the form of the lowest order 
Lagrangian is
\beq
{\cal L}_{\phi B}^{W \, (0)}  =  \:
d \, \tr \Big( \bar{B}  \{ h_+ , B\} \Big) + \:
f \, \tr \Big( \bar{B}  [ h_+ , B ] \Big) \qquad ,
\eeq
and $d,f$ are the {\it only} weak counterterms considered in most previous 
calculations \cite{DGH,BSW,J}. As discussed in the introduction,
use of this Lagrangian does not provide a simultaneously 
satisfactory fit to
s-- and p--waves, even after the inclusion of meson loops.
In order to improve the agreement with experiment, one must account for
additional weak
counterterms, but, performing the calculation with the complete Lagrangian 
including counterterms from higher orders,
one has no predictive power\cite{BH}. 
Indeed there exist only eight experimental numbers:
the s- and p-wave amplitudes for the four independent hyperon decays, while
on the other side, the theoretical predictions contain considerably more 
than eight low energy constants.
We are not able to fix all the low--energy
constants appearing in ${\cal L}_{\phi B}^W$
from data, even if we resort to large $N_c$ arguments.
We will therefore use the principle of resonance saturation in order to
estimate the importance of these constants,
as outlined in the following sections.

\subsection{Inclusion of resonances}
In order to include resonances one begins by writing down the most
general Lagrangian at lowest order which exhibits the same symmetries
as the underlying theory, {\it i.e.} Lorentz invariance and chiral symmetry.
For the strong part we require invariance under $C$ and $P$ transformations
separately, while
the weak piece is invariant under $CPS$ transformations,
where the transformation $S$
interchanges down and strange quarks in the Lagrangian.
We will work in the $CP$ limit so that all LECs are real.
(Of course, $C$ and $P$ invariance 
are not required separately for the weak interacting Lagrangian.)

We begin with the inclusion of the lowest --- 1/2$^-$ ---
negative parity level in (70,$1^-$).
In \cite{LeY} it was shown that such states dominate the contributions
from (70,$1^-$) and considerably improve agreement with experiment for the
hyperon decays. 
Among the well
established states one has $N(1535)$ and $\Lambda(1405)$. As for the remaining
1/2$^-$ octet components there are a number of not so well--established
states in the same mass range --- cf. \cite{LeY} and references therein.
We denote the 1/2$^-$ octet by $R$. 
Under $CP$ transformations the fields behave as
\beqa
B \q & \rightarrow & \q \gamma_0 C \bar{B}^{T} \q , \q
\bar{B} \q \rightarrow  \q B^{T} C \gamma_0  \q , \q
u^{\mu} \q \rightarrow  \q - u_{\mu}^{T} \q , \q \no \\
h_+ \q & \rightarrow &  \q h_+^{T} \qq , \qq
D^{\mu} \q \rightarrow  \q - D_{\mu}^{T} \q , \q \no \\
R \q & \rightarrow & \q - \gamma_0 C \bar{R}^{T} \q , \q
\bar{R} \q \rightarrow  \q - R^{T} C \gamma_0  \q , \q
\eeqa
where $C$  is the usual charge conjugation matrix.
The kinetic term is straightforward
\beq
{\cal L}_{R}^{kin}  
 = 
 i \, \tr \Big( \bar{R} \gamma_{\mu} [  D^{\mu} , R] \Big) -
M_R \, \tr \Big( \bar{R} R \Big)
\eeq
with $M_R$ being the mass of the resonance octet in the chiral limit.
The resonances R couple strongly to the 1/2$^+$ baryon octet $B$ via the
Lagrangian
\beq
{\cal L}_{R B}^{(1)}  = 
i s_d \Big[ \, \tr \Big( \bar{R} \gamma_{\mu} \{ u^{\mu}, B\} \Big)
       - \tr \Big( \bar{B} \gamma_{\mu} \{ u^{\mu}, R\} \Big) \: \Big] +
 i s_f \Big[ \, \tr \Big( \bar{R} \gamma_{\mu} [ u^{\mu}, B] \Big)
       - \tr \Big( \bar{B} \gamma_{\mu} [ u^{\mu}, R] \Big) \: \Big] 
\eeq
and the two coupling constants $s_d$ and $s_f$ 
can be determined from the strong
decays of the resonances (cf. App.~\ref{app:a}), yielding the central 
values
\beq
s_d = 0.17  \qq , \qq s_f = \, - \, 0.12  \qq .
\eeq
A few remarks about the Lagrangian ${\cal L}_{R B}^{(1)}$ are in order.
In principle, terms of the form $\bar{B} u^{\mu} D_{\mu} R $
are allowed by symmetry considerations. But, by use of the equation of motion
for the resonance fields
\beq
 i \, \gamma_{\mu} [  D^{\mu} , R]  -M_R \, R \: = \: 0
\eeq
one is able to reduce it to the terms already included in 
${\cal L}_{R B}^{(1)}$.
The interaction term $ i \bar{R} \gamma_5 B $ also satisfies the
symmetry constraints, but can be transformed away by a unitary
transformation. The proof of this is as follows.
Consider a Lagrangian of the form
\beq
{\cal L}  =   i \bar{B} \gamma_\mu D^\mu B - M_B \bar{B} B
               +i \bar{R} \gamma_\mu D^\mu R - M_R \bar{R} R   
  + i \alpha \bar{R} \gamma_5 B + i \alpha \bar{B} \gamma_5 R
\eeq
where we have supressed flavor indices and $\alpha$ is the off diagonal
coupling. 
By introducing a doublet notation we can rewrite the Lagrangian
\beq
{\cal L}  =   i \bar{Q} \gamma_\mu D^\mu Q +  \bar{Q} ( A + i \gamma_5 C) Q
\eeq
with
\beqa
Q = \left( \matrix{ B \no \\ R \no \\}
\!\!\!\!\!\!\!\!\!\!\!\!\!\!\!\! \right) \qq , \qq \q 
A = \left( \matrix{ -M_B & 0 \no \\ 0 & -M_R \no \\}
\!\!\!\!\!\!\!\!\!\!\!\!\!\!\!\! \right)  \qq , \qq \q 
C = \left( \matrix{ 0 & \alpha \no \\ \alpha & 0 \no \\}
\!\!\!\!\!\!\!\!\!\!\!\!\!\!\!\! \right)  \qq . 
\eeqa
Note, that $A$ and $C$ are hermitian.
Decomposing the doublet field $Q$ as follows
\beq
Q_R = \frac{1}{2} ( 1 + \gamma_5 ) Q  \qq , \qq
Q_L = \frac{1}{2} ( 1 - \gamma_5 ) Q  
\eeq
the Lagrangian reads
\beq \label{lag}
{\cal L}  =   i \bar{Q}_R \gamma_\mu D^\mu Q_R +
 i \bar{Q}_L \gamma_\mu D^\mu Q_L + \bar{Q}_L M Q_R + \bar{Q}_R M^\dagger Q_L
\eeq
with $M = A + iC$.
Then by applying a bi--unitary transformation
\beq
Q_R \rightarrow R Q_R \qq , \qq
Q_L \rightarrow L Q_L
\eeq
with unitary matrices $R$ and $L$ one can diagonalize the matrix $M$
\beq
L^\dagger M R = M_d
\eeq
where $M_d$ is diagonal with positive elements.
The first two terms in Eq.~(\ref{lag}) remain unchanged by this transformation.
Expressing the Lagrangian in terms of $Q$ we obtain
\beq
{\cal L}  =   i \bar{Q} \gamma_\mu D^\mu Q +  \bar{Q} M_d Q
\eeq
which is the desired result.
Including the interaction terms from ${\cal L}_{R B}^{(1)}$ does not
alter the proof.
The interaction term of the form $ i \bar{R} \gamma_5 B $ can therefore
be neglected, which leads to significant simplifications of the weak
Lagrangian between the resonances and the low--lying baryon octet.

We can then turn to the lowest order weak Lagrangian which reads
\beq
{\cal L}_{R B}^{W \,(1)}  = 
i w_d \Big[ \, \tr \Big( \bar{R}  \{h_+ , B\} \Big)
       - \tr \Big( \bar{B}  \{ h_+, R\} \Big) \: \Big] +
  i w_f \Big[ \, \tr \Big( \bar{R}  [h_+ , B] \Big)
       - \tr \Big( \bar{B}  [ h_+, R] \Big) \: \Big] \q .
\eeq
with two unknown weak couplings $w_d$ and $w_f$, which will be determined 
from a fit to the hyperon decays.
(Again, a term of the form $\bar{R} \gamma_5 h_+ B$ is allowed by
symmetry considerations, but a proof analogous to the one above shows
that such terms can be transformed away.)
Furthermore, terms with the structure
$i \bar{R} \gamma_\mu h_+ u^\mu B$ and
$ \bar{R} \gamma_\mu \gamma_5 (D^\mu h_+)  B$ 
are possible. But, after contraction with the vertices from
${\cal L}_{R B}^{(1)}$ in the resonance saturation scheme, they
deliver contact terms of chiral order two and involve
at least two outgoing mesons,
which is clearly beyond our tree level considerations.

We will not include any additional resonances from the (70,1$^-$) multiplet,
which were the only states considered in \cite{LeY}. But in many other
applications the spin--3/2$^+$ decuplet and the spin--1/2$^+$ Roper octet
play an important role, cf. {\it e.g.} \cite{BM}. The decuplet is only 
231 MeV higher in  average than the ground state octet and 
the Roper octet masses are comparable to the 1/2$^-$ states $R$.
One should therefore presumably account for these resonances.

We first consider the decuplet.
Due to angular momentum conservation
the spin--3/2 decuplet states can couple to the spin--1/2 baryon octet
only accompanied by Goldstone bosons --- {\it i.e.}
decuplet states contribute only through
loop diagrams to non--leptonic hyperon decays.
An explicit calculation shows that such diagrams saturate
contact terms of the same chiral order --- {\it O}($p^2$) ---
as the loop corrections 
with the baryon octet \cite{BH}.
Since we restrict ourselves to {\it O}($p^0$) and 
{\it O}($p^1$) we can disregard
such decuplet contributions. 
In addition, the calculation of relativistic loop diagrams in the
resonance saturation scheme leads to some complications. 
The integrals are in 
general divergent and have to be renormalized, which introduces
new unknown parameters. 
The absence of a strict chiral counting scheme in the relativistic 
formulation leads to contributions from higher loop diagrams
which are usually neglected in such calculations, cf. \cite{BM}.

The lowest multiplet of excited states contributing to
the chiral order {\it O}($p^1$)
is the octet of even--parity
Roper--like spin--1/2 fields. While it was argued in \cite{JM} that
these play no role, a more recent study seems to indicate that one
cannot neglect contributions from these states to,
{\it e.g.}, the decuplet magnetic moments \cite{BaM}. It is thus
important to investigate the possible contribution of these baryon
resonances to the LECs.
The octet consists of the $N^*(1440)$, $\Sigma^*(1660)$,  
$\Lambda^*(1600)$ and $\Xi^*(1620?)$.
We denote the spin--1/2$^+$ resonance octet by $B^*$.
The transformation properties of $B^*$ under $CP$ are the same as for
the ground state baryons $B$, and
the effective Lagrangian of the $B^*$ octet coupled to the ground state
baryons takes the form
\beq
{\cal L}_{B^* B} = {\cal L}_{B^*}^{kin} + 
                   {\cal L}_{B^* B}^S + {\cal L}_{B^* B}^W
\eeq
with the kinetic term
\beq
{\cal L}_{B^*}^{kin} =
 i \, \tr \Big( \bar{B}^* \gamma_{\mu} [  D^{\mu} , B^*] \Big) -
M_{B^*} \, \tr \Big( \bar{B}^* B^* \Big) \q ,
\eeq
a strong interaction part \cite{BM}
\beqa
{\cal L}_{B^* B}^S  &=& 
\frac{1}{4} D^* \Big[ \, \tr \Big( \bar{B^*} \gamma_{\mu} \gamma_5
          \{ u^{\mu}, B\} \Big)
 + \tr \Big( \bar{B} \gamma_{\mu} \gamma_5 \{ u^{\mu}, B^*\} \Big) \: \Big]
\no \\ &+&
\frac{1}{4} F^*  \Big[ \, \tr \Big( \bar{B^*} \gamma_{\mu} \gamma_5
   [ u^{\mu}, B] \Big)
 + \tr \Big( \bar{B} \gamma_{\mu} \gamma_5 [ u^{\mu}, B^*]\Big)\: \Big] \q .
\eeqa
and a weak piece
\beq
{\cal L}_{B^* B}^{W}  = 
d^* \Big[ \, \tr \Big( \bar{B}^*  \{h_+ , B\} \Big)
       + \tr \Big( \bar{B}  \{ h_+, B^* \} \Big) \: \Big] +
 f^* \Big[ \, \tr \Big( \bar{B}^*  [h_+ , B] \Big)
       + \tr \Big( \bar{B}  [ h_+, B^*] \Big) \: \Big] \q .
\eeq
The couplings $D^*$ and $F^*$ have already been determined from the
strong decays of these resonances \cite{BM}, with central values
\beq
D^* = 0.60 \qq ,\qq F^* = 0.11 \qq,
\eeq
while the weak parameters $d^*$ and $f^*$ can be determined from a fit to the 
non--leptonic hyperon decays --- cf. Sec.~3.

Having written down the relevant Lagrangian for the resonances coupled
to the ground state baryons we can proceed to integrate out
the heavy degrees of freedom from the effective theory.

\subsection{Resonance saturation}
In this section we calculate the tree level diagrams involving
resonances which contribute to non--leptonic hyperon decay. Allowing
the resonance masses to become large with fixed ratios of coupling
constant to mass, higher order terms in the effective meson--baryon
Lagrangian are generated, the coefficients of which can be expressed in terms
of a few resonance parameters.

Using the vertices from the Lagrangians developed in the preceeding section we
calculate the diagrams in Fig.~1. 
Then, performing the limit $M_R, M_{B^*}
\rightarrow \infty$ and using the Cayley--Hamilton identity for
the two traceless $3 \times 3$ matrices $u_\mu$ and $h_+$
\beqa
& & \frac{3}{2} \tr  \Big( \bar{B} \Gamma_\mu \{h_+, \{u^\mu , B \} \} \Big)
+ \frac{3}{2} \tr  \Big( \bar{B} \Gamma_\mu \{u^\mu, \{h_+ , B \} \} \Big) 
\no \\
& & +\frac{1}{2} \tr \Big( \bar{B} \Gamma_\mu [h_+, [u^\mu , B ]] \Big)
+\frac{1}{2} \tr \Big( \bar{B} \Gamma_\mu [u^\mu, [ h_+ , B ] ] \Big) \no \\
& = &  2 \tr \Big( \bar{B} \Gamma_\mu B \Big)  \tr \Big( h_+ u^\mu \Big)
+ 2 \tr \Big( \bar{B} h_+ \Big) \Gamma_\mu  \tr \Big(  u^\mu B \Big)
+ 2 \tr \Big( \bar{B} u^\mu \Big) \Gamma_\mu  \tr \Big(  h_+ B \Big) 
\eeqa
with $\Gamma_\mu = \gamma_\mu \gamma_5 , \gamma_\mu $,
one generates the effective Lagrangian
\beqa
{\cal L}_{\phi B}^{W \, (1)}
& = & 
g_{1} \bigg\{ \tr \Big( \bar{B} \gamma_{\mu}    
                            [ h_+ , \{ u^{\mu}, B\} ] \Big) +
  \tr \Big( \bar{B} \gamma_{\mu}  
                      \{ u^{\mu} , [h_+, B] \} \Big) \bigg\} \no \\
& + & 
g_{2} \bigg\{ \tr \Big( \bar{B} \gamma_{\mu}    
                               \{ h_+ , [u^{\mu}, B] \} \Big) +
  \tr \Big( \bar{B}\gamma_{\mu}  
                          [ u^{\mu} , \{ h_+, B\} ] \Big) \bigg\} \no \\
& + & 
g_{3} \bigg\{ \tr \Big( \bar{B} \gamma_{\mu}  
                             [ h_+ , [u^{\mu}, B] ] \Big) +
     \tr \Big( \bar{B} \gamma_{\mu}  
                         [ u^{\mu} , [h_+, B] ] \Big) \bigg\} +
g_{4} \tr \Big( \bar{B} \gamma_{\mu}    B \Big) 
                             \tr \Big( u^{\mu} h_+ \Big) \no \\
& + &
g_{6} \bigg\{ \tr \Big( \bar{B} \gamma_{\mu}  \gamma_5  
                            [ h_+ , \{ u^{\mu}, B\} ] \Big) +
  \tr \Big( \bar{B} \gamma_{\mu}  \gamma_5  
                      \{ u^{\mu} , [h_+, B] \} \Big) \bigg\} \no \\
& + & 
g_{7} \bigg\{ \tr \Big( \bar{B} \gamma_{\mu}    \gamma_5  
                               \{ h_+ , [u^{\mu}, B] \} \Big) +
  \tr \Big( \bar{B}\gamma_{\mu}  \gamma_5  
                          [ u^{\mu} , \{ h_+, B\} ] \Big) \bigg\} \no \\
& + & 
g_{8} \bigg\{ \tr \Big( \bar{B} \gamma_{\mu}  \gamma_5  
                             [ h_+ , [u^{\mu}, B] ] \Big) +
     \tr \Big( \bar{B} \gamma_{\mu}  \gamma_5  
                         [ u^{\mu} , [h_+, B] ] \Big) \bigg\} \no \\
& + & 
g_{9} \tr \Big( \bar{B} \gamma_{\mu}  \gamma_5    B \Big) 
                             \tr \Big( u^{\mu} h_+ \Big) \qq .
\eeqa
The couplings read, in terms of the resonance parameters,
\beqa
g_{1} & = &  \frac{s_d \, w_f}{M_R} \q , \q
g_{2}   =    \frac{s_f \, w_d}{M_R} \q , \q
g_{3}   =    \frac{s_f \, w_f}{M_R} - \frac{s_d \, w_d}{3 M_R} \q , \q
g_{4}   =    \frac{4 s_d \, w_d}{3 M_R} \q , \q \no \\
g_{6} & = &  \frac{D^* \, f^*}{4 M_{B^*}} \q , \q
g_{7}   =    \frac{F^* \, d^*}{4 M_{B^*}} \q , \q
g_{8}   =    \frac{F^* \, f^*}{4 M_{B^*}} -\frac{D^*\, d^*}{12M_{B^*}}\q , \q
g_{9}   =    \frac{D^* \, d^*}{3 M_{B^*}} \q .
\eeqa
The Lagrangian ${\cal L}_{\phi B}^{W \, (1)}$ forms, together with
the weak Lagrangians ${\cal L}_{\phi B}^{W \, (0)}$ at lowest order and
${\cal L}_{\phi }^{W}$ from Eq.~(\ref{weak}), the strangeness changing
Lagrangian which we employ for the calculation of the decay amplitudes.
(Note that the most general Lagrangian ${\cal L}_{\phi B}^{W \, (1)}$ 
contains two additional terms
\beq
g_{5} \, \tr \Big( \bar{B} h_+ \Big) \gamma_{\mu}  \tr \Big( u^{\mu} B \Big)
\q + \q
g_{10}\,  \tr \Big( \bar{B} h_+ \Big) \gamma_{\mu} \gamma_5 \tr 
\Big( u^{\mu} B \Big)
 \q + \q  (h.c.)
\eeq
that are not generated by the resonances considered here.)

\subsection{Heavy baryon limit}
We evaluate the decay amplitudes in the extreme non--relativistic limit
wherein the baryons are characterized
by a four velocity $v_\mu$ \cite{JM1}. A consistent chiral counting
scheme emerges, {\it i.e.} a one--to--one correspondence between
the Goldstone boson loops and the expansion in small momenta and
quark masses.
In the heavy baryon formulation the p-wave must be modified, since 
$\gamma_5$ connects the light with the heavy degrees of freedom which are 
integrated out in this scheme.
One therefore introduces the modified heavy baryon p-wave amplitude
${\cal A}_{ij}^{(P)} $ by
\beq
A_{ij}^{(P)} = - \frac{1}{2} (E_j + M_j) {\cal A}_{ij}^{(P)} \qq ,
\eeq
where $E_j$ and $ M_j$ are the energy and mass of the outgoing baryon, 
respectively.
In the rest frame of the heavy baryon, $ v_{\mu} = (1,0,0,0) $,
the decay amplitude reduces to the non--relativistic form
\beqa
{\cal A}( B_i \rightarrow B_j \, \pi) & = &
\bar{\chi}_{B_j} \, \Big\{ \, {\cal A}_{ij}^{(S)} + \, 
\frac{1}{2}\,  {\bf k} \cdot {\bf \sigma}\,  {\cal A}_{ij}^{(P)}\, 
\Big\}\chi_{B_i}\no \\
& = &
\bar{\chi}_{B_j} \, \Big\{ \, {\cal A}_{ij}^{(S)} + \, 
S \cdot k \, {\cal A}_{ij}^{(P)}\,  \Big\}\chi_{B_i} \qq ,
\eeqa
where $k$ is the outgoing momentum of the pion and $2 S_\mu = i \gamma_5
\sigma_{\mu \nu} v^\nu$
is the Pauli-Lubanski spin vector, which
in the rest frame is given by
$S_{\mu}^{{\bf v}=0} =  ( 0 , \frac{1}{2} {\bf \sigma} ) $.
The structure of the Lagrangian remains almost unchanged with
$\gamma_\mu$ and $\gamma_\mu \gamma_5$ replaced by
$v_\mu$ and $2 S_\mu$, respectively.
The only additional terms that contribute are relativistic corrections to the
Dirac term and are of the form \cite{BH}
\beq \label{cor}
{\cal L}_{\phi B}^{(2)}   =  {\cal L}_{\phi B}^{(2,rc)} 
 = 
- \frac{1}{2 \mnod} \tr \Big( \bar{B} [ D_{\mu}, [D^{\mu}, B]] \Big) 
+ \frac{1}{2 \mnod} \tr \Big( \bar{B} [ v \cdot D, [v \cdot D, B]] \Big) 
\eeq
We utilize the same notation for the baryon fields as in the relativistic case.
These terms produce a finite shift to the bare masses of the baryons.
Since we work with the physical masses of the baryons the effects
of ${\cal L}_{\phi B}^{(2)}$ are already included in our expressions
for the decay amplitudes and we can neglect (\ref{cor}).

The general structure of the s-wave decay amplitudes is
\beqa   \label{swa}
{\cal A}_{ij}^{(s)} & = &
\frac{1}{\sqrt{2}\, F_{\pi}}\, \bigg\{ \,
\alpha_{ij}^{(s)} \, + \,
v \cdot k \: \beta_{ij}^{(s)} \bigg\}
\eeqa
with
\beqa
\alpha_{\Sigma^+ n}^{(s)}  & = &  0  \qq  \qq \qq \q \; \:  \qq 
\beta_{\Sigma^+ n}^{(s)}   =   -4 g_1 + 4 g_2 - 4 g_3   \no \\ 
\alpha_{\Sigma^- n}^{(s)} &  =&    d -f  \qq \qq \qq \! \!  \qq 
\beta_{\Sigma^- n}^{(s)}   =   -2 g_1 - 2 g_2 + 2 g_3   \no \\ 
\alpha_{\Lambda p }^{(s)}  & = &   - \frac{1}{\sqrt{6}} (d+3f) \qq \qq 
\beta_{\Lambda p }^{(s)}  =    - \frac{1}{\sqrt{6}} 
                    ( 10 g_1 + 2 g_2 + 6 g_3 )  \no \\ 
\alpha_{\Xi^- \Lambda }^{(s)}  & = &   - \frac{1}{\sqrt{6}} (d-3f) \qq \qq 
\beta_{\Xi^- \Lambda }^{(s)} =     \frac{1}{\sqrt{6}} 
                    ( 10 g_1 + 2 g_2 - 6 g_3 )   \qq .
\eeqa
In the rest frame of the decaying baryon the energy
of the meson may be written as
\beq \label{vq}
v \, \cdot \, k = \frac{1}{2 \, M_i} 
             \Big( M_i^2 - M_j^2 + m_{\pi}^2 \Big) \qq .
\eeq 
We obtain very similar expressions for the resonance contributions
to those found in the constituent quark model \cite{LeY}. 
There the contributions
to the s--waves were found to be proportional to the mass difference
of the decaying and the light baryon which differs from $v \cdot k$
only by terms quadratic in the meson masses.
To the order we are working then we have agreement with the quark model
calculation.
The contact diagram that contributes to the s-waves is shown in Fig.~2a.

For the p--waves one finds the form
\beqa   \label{pwa}
{\cal A}_{ij}^{(p)} & = &
\frac{1}{\sqrt{2}\, F_{\pi}}\, \bigg\{ \,
\alpha_{ij}^{(p)} \, + \, \beta_{ij}^{(p)} 
+ \frac{1}{2} h_\pi \frac{m_\pi^2}{m_\pi^2 - m_K^2} \phi_{ij}^{(p)}
\bigg\}
\eeqa
where $\alpha_{ij}^{(p)}$ denotes the baryon pole terms, $\beta_{ij}^{(p)}$
the contact terms in ${\cal L}_{\phi B}^{W (1)} $
and $\phi_{ij}^{(p)}$ the contribution from the weak decay of the meson.
The diagrams which contribute to the p--waves are depicted in Fig.~2,
and yield
\beqa
\alpha_{\Sigma^+ n}^{(p)} & = &
    - \frac{1}{M_\Sigma - M_N} 2 \, D \, (d-f)
  - \frac{1}{M_\Lambda - M_N} \frac{2}{3} \, D \, (d+3f) \no \\
\beta_{\Sigma^+ n}^{(p)}  & = &   -8 g_6 + 8 g_7 - 8 g_8  \q , \q
\phi_{\Sigma^+ n}^{(p)} = 0 \no \\  
\alpha_{\Sigma^- n}^{(p)} & = &  
- \frac{1}{M_\Sigma - M_N} 2 \, F \, (d-f) 
- \frac{1}{M_\Lambda - M_N} \frac{2}{3} \, D \, (d+3f) \no \\
\beta_{\Sigma^- n}^{(p)}  & = &  -4 g_6 - 4 g_7 + 4 g_8  \q , \q
\phi_{\Sigma^- n}^{(p)} = D - F \no \\   
\alpha_{\Lambda p}^{(p)} & = &  
\frac{1}{M_\Lambda-M_N} \frac{2}{\sqrt{6}} (d+3f)(D+F) \,
+ \frac{1}{M_\Sigma - M_N} \frac{4}{\sqrt{6}}\, D\,(d-f) \no \\
\beta_{\Lambda p }^{(p)} & = &    - \frac{1}{\sqrt{6}} 
                    ( 20 g_6 + 4 g_7 + 12 g_8 )  \q , \q
\phi_{\Lambda p}^{(p)} = - \frac{1}{\sqrt{6}} ( D +3F ) \no \\  
\alpha_{\Xi^- \Lambda}^{(p)} & = &  
-\frac{1}{M_\Xi-M_\Lambda} \frac{2}{\sqrt{6}} (d-3f)(D-F) \,
- \frac{1}{M_\Xi-M_\Sigma} \frac{4}{\sqrt{6}}\, D\,(d+f) \no \\
\beta_{\Xi^- \Lambda}^{(p)} & = & \frac{1}{\sqrt{6}} 
                    ( 20 g_6 + 4 g_7 - 12 g_8 ) \q , \q
\phi_{\Xi^- \Lambda}^{(p)} = - \frac{1}{\sqrt{6}} ( D -3F )  \qq .
\eeqa

\section{Results and discussion} \label{sec:dis}

In this section we discuss the numerical values of the LECs and the 
fit to experiment. There exist eight independent
experimental numbers, {\it i.e.} s-- and 
p--wave amplitudes for the four decays 
$\Sigma^+  \rightarrow n \, \pi^+ \, , \, 
\Sigma^- \rightarrow n \, \pi^- \, , \, 
\Lambda \rightarrow p \, \pi^- \, \mbox{and} \,
\Xi^- \rightarrow  \Lambda \, \pi^- $,
which are not related by isospin. The central values for our parameters
are $ F_{\pi} = 92.4 $ MeV, $ D= 0.75$, $F=0.50$.
The procedure of estimating the counterterms of higher order
in the resonance saturation scheme 
involving the 1/2$^-$, 1/2$^+$ octets $R, B^*$
introduces eight additional parameters,
four of which are determined from the strong resonance decays.
Together with the couplings $d$ and $f$ from the weak Lagrangian
at lowest order we have then six parameters 
with which to perform a least--squares fit
to the non--leptonic hyperon decays.
The experimental values of the decays are shown in Table~1,
and the fitted 
chiral expansions of the decay 
amplitudes read in units of $10^{-7}$
\beqa
{\cal A}_{\Sigma^+ n}^{(s)} & = & 0.00 -  0.04  =  - 0.04 \q ,\q
{\cal A}_{\Sigma^+ n}^{(p)}   =   - 19.6 - 24.8 =  -44.4 \q ,\q \no \\
{\cal A}_{\Sigma^- n}^{(s)} & = & 7.19 - 1.86  = 5.33 \q ,\q
{\cal A}_{\Sigma^- n}^{(p)}   =  - 5.33 + 6.94 = 1.61 \q ,\q \no \\
{\cal A}_{\Lambda p}^{(s)} & = &   3.32 - 1.15  = 2.17 \q ,\q
{\cal A}_{\Lambda p}^{(p)}   =  - 12.4 - 11.0 = -23.4 \q ,\q \no \\
{\cal A}_{\Xi^- \Lambda}^{(s)} & = & -6.06 + 1.92 = -4.14 \q ,\q
{\cal A}_{\Xi^- \Lambda}^{(p)}   =  -10.44 - 4.30 = -14.7 \q ,
\eeqa
where the first number is the lowest order piece,
involving the weak counterterms $d,f$,  and the second
number contains the contributions from contact terms at next order.
The contributions from the weak meson decays for the p--waves appear
also at that order and are, therefore, included in the second number.
The results, particularly for the p--waves, are in satisfactory agreement with
experiment.
A fit with {\it only} the spin--1/2$^-$ resonances, as performed in the quark
model in \cite{LeY}, does {\it not} lead to good agreement between theory
and experiment in the framework of heavy baryon chiral perturbation theory,
although the 1/2$^-$ contribution to the parity--violating decay 
amplitudes is roughly of the same size as in \cite{LeY}. 
The reason for this is that it is not possible 
to obtain a satisfactory fit for p--waves by using
only the lowest order couplings $d$ and $f$. 
In the usual quark model approach
the p--wave amplitudes include explicit
$SU(3)$ symmetry breaking corrections of higher chiral order. 
A much improved fit to the p--waves is possible and the contributions
from the spin--1/2$^-$ resonances, which contribute only to the s--waves,
are sufficient to achieve a satisfactory fit for both s-- and p--waves.
This indicates that such higher order
corrections are essential. By including the
spin--1/2$^+$ resonances in the resonance saturation scheme,
which contribute to the p--waves at next--to--leading order, one is
able to account for some of these higher order effects.
The contributions from the 1/2$^+$ resonances 
to the parity--conserving decay amplitudes are of comparable size as the
the ground state contributions, and apparently, these resonances are crucial
in heavy baryon chiral perturbation theory
to obtain a satisfactory fit also for the p--waves.
For completeness, the corresponding values of the couplings
$g_i$ from the Lagrangian ${\cal L}_{\phi B}^{W (1)}$
are given in Table~2.

It should be noted that
a very different fit with six parameters was performed in \cite{BH}.
There loop corrections were included and the exchange
of intermediate decuplet states was used in order to estimate the LECs.
It turned out that no satisfactory fit was possible and additional counterterms
had to be included. Inclusion of the 1/2$^-$ and 1/2$^+$ 
resonance states seems then to play an important role
for understanding of the non--leptonic hyperon decays. 
In fact in the case of p--waves
their contribution is comparable to or even exceeds that from
lowest order pieces. In order to make a more definitive
statement about their importance one must, of course, go to higher orders
and include loops. However, this is beyond the scope of the present work.

\section{Summary}

We have in this paper 
studied the importance of baryon resonances for the non--leptonic 
hyperon decays at tree level in chiral perturbation theory.
To this end, we included the spin--1/2$^-$ octet from the (70,1$^-$) states
and the octet of Roper--like 1/2$^+$ fields in the
effective theory. 
The most general Lagrangian incorporating these resonances coupled to the 
ground state baryons introduces eight new parameters, four of which can be
determined from the strong decays of the resonances.
Integrating out the resonances generates counterterms
in the Lagrangian at next--to--leading order. On the other hand, 
the inclusion of the spin--3/2$^+$ decuplet, as performed
in \cite{BH}, generates terms at the
same chiral order as the loop corrections --- {\cal O}($p^2$) ---, 
which is beyond the accuracy
of this calculation and therefore can be neglected.
In \cite{LeY} it was argued that the inclusion of the spin--1/2$^-$ octet
is sufficient to obtain a satisfactory fit for both s-- and p--waves.
We were able to show that in the framework of chiral
perturbation theory the structure of the contributions from these resonances
agrees with the results in the quark model to the order we are working. 
In the quark model the expressions for the p--waves include additional explicit
$SU(3)$ symmetry breaking corrections of second chiral order, in which case,
a much improved fit to the p--waves is possible and the contributions
from the spin--1/2$^-$ resonances, which contribute only to the s--waves,
are sufficient to achieve a satisfactory fit for both s-- and p--waves.
On the other hand, in chiral perturbation theory
the improvement of experimental agreement is brought about by
the inclusion of the Roper--octet, which is in the same mass
range as the 1/2$^-$ octet. The reason for this is that
the contributions from the lowest order couplings $d$ and $f$ 
for the p--wave decay amplitudes tend to cancel thus enhancing the 
contributions from terms of higher chiral order.
By including the
spin--1/2$^+$ resonances in the resonance saturation scheme,
which contribute to the p--waves at next--to--leading order, one is
able to overcome this problem.
By fitting the six parameters of the weak Lagrangian, two from
lowest order and four at next--to--leading order, to the
eight independent decay amplitudes that are not related by isospin,
we obtain satisfactory agreement with experiment.
We suggest then that the inclusion of spin--1/2 resonances in non--leptonic
hyperon decays provides a reasonable estimate
of the importance of higher order 
counterterms. In order to make a more definite statement,
one should go to higher orders and include meson loops.

Of course, our fit is not unique.
Another satisfactory fit for the decay amplitudes was achieved in \cite{BH}
by including Goldstone boson loops and spin--3/2 decuplet contributions.
The effects of higher resonances like the ones considered in the present work
were neglected.
By considering only the non--leptonic hyperon
decays it is not possible to decide
which approach describes nature more appropriately. One must examine
other weak processes involving hyperons, {\it e.g.} the radiative hyperon 
decays. This work is under way \cite{BH2}.

\section*{Acknowledgements}
This work was supported in part by the Deutsche Forschungsgemeinschaft
and by the National Science Foundation.

\appendix 
\def\theequation{\Alph{section}.\arabic{equation}}
\setcounter{equation}{0}
\section{Determination of the $\frac{1}{2}^-$--resonance couplings $s_d$ and
         $s_f$} \label{app:a}
The decays listed in the particle data book,
which determine the coupling constants $s_d$ and $s_f$,
are $N(1535) \rightarrow N \pi$ , $N(1535) \rightarrow N \eta$ and
$\Lambda(1405) \rightarrow \Sigma \pi$.
The width follows via
\beq
\Gamma = \frac{1}{8 \pi M_R^2} |{\bf k}_\phi| |{\cal T}|^2
\eeq
with
\beq
|{\bf k}_\phi| = \frac{1}{2 M_R} \Big[ (M_R^2-(M_B+m_\phi)^2)
                 \, (M_R^2-(M_B-m_\phi)^2) \Big]^{1/2}
\eeq
being the three--momentum of the meson $\phi = \pi , \eta$
in the rest frame of the resonance.
The terms $M_R$ and $M_B$ are the masses of the resonance and the
ground state baryon, respectively.
For the transition matrix one obtains
\beq
|{\cal T}|^2 = \frac{2}{F_\pi^2}  ( M_R - M_B)^2 
     \Big[ (M_R + M_B)^2 - m_\phi^2 \Big] A_{R \phi}
\eeq
with the coefficients
\beq
A_{N(1535) \, \pi} = \frac{3}{2} ( s_d + s_f )^2 \qq , \qq
A_{N(1535) \, \eta} = \frac{1}{6} ( s_d - 3 s_f )^2 \qq , \qq
A_{N(1535) \, \pi} = 2 s_d^2 \q .
\eeq
Using the experimental values for the decay widths
we arrive at the central values
\beq
s_d = 0.17 \qq , \qq s_f = - \, 0.12 
\eeq
where we have chosen the sign of $s_d$ to be positive in accordance
with the ground state octet $D$ coupling.
We do not present the uncertainties in these parameters here, since
for the purpose of our considerations a rough estimate of these constants
is sufficient.

\newpage

\section*{Table captions}

\begin{enumerate}

\item[Table 1] Experimental values of the decay amplitudes 
           including the errors.
           The numbers have to be multiplied by a factor of $10^{-7}$.

\item[Table 2] Numerical values of the LECs obtained from a 
           least--squares fit. The couplings $d$ and $f$ are given
          in units of $10^{-7}$ GeV, the $g_i$ in units of $10^{-7}$.

\end{enumerate}

\vskip 1.2in


\section*{Figure captions}

\begin{enumerate}

\item[Fig.1] Diagrams including resonances that contribute to s-- and
             p-- waves. Solid and dashed lines denote ground state baryons
             and Goldstone bosons, respectively. The double line represents
             a resonance. Solid squares and circles are vertices of the
             weak and strong interactions, respectively.

\item[Fig.2] Diagrams that contribute to s-- and
             p-- waves. Fig.~2a contributes both to s-- and p--waves,
             whereas Figures 2b,c,d contribute only to the p--waves.
             Solid and dashed lines denote ground state baryons
             and Goldstone bosons, respectively.
             Solid squares and circles are vertices of the
             weak and strong interactions, respectively.

\end{enumerate}

\newpage

\begin{center}

\begin{table}[bht] 
\begin{center}
  \begin{tabular}{|cccc|}
    \hline
     ${\cal A}_{\Sigma^+ n}^{(s)}$ & 
      ${\cal A}_{\Sigma^- n}^{(s)}$ & 
      ${\cal A}_{\Lambda p}^{(s)}$ &  
      ${\cal A}_{\Xi^-\Lambda }^{(s)}$ 
        \\
    \hline
      0.13 $\pm$ 0.02 & 4.27 $\pm$ 0.02&  
          3.25 $\pm$ 0.02 &
         $-$4.51 $\pm$ 0.02  \\
      \hline
       \hline
     ${\cal A}_{\Sigma^+ n}^{(p)}$ & ${\cal A}_{\Sigma^- n}^{(p)}$ & 
      ${\cal A}_{\Lambda p}^{(p)}$ &  ${\cal A}_{\Xi^-\Lambda}^{(p)}$ \\
    \hline
     $-$44.4 $\pm$ 0.16 &   1.52 $\pm$ 0.16  &  $-$23.4  $\pm$ 0.56
      & $-$14.8 $\pm$ 0.55 \\
      \hline
  \end{tabular}
  \medskip
\end{center}
\end{table}

\vskip 0.7cm

Table  1

\vskip 1.5cm
\begin{table}[bht]\label{table2}
\begin{center}
  \begin{tabular}{|cccccccccc|}
    \hline
      $d$ & $f$ & $g_1$  
      & $g_2$ &  $g_3$ & $g_4$ & $g_6$ & $g_7$ & $g_8$ & $g_9$ \\
    \hline
 0.44 & $-$0.50 & 0.26 &0.14 & $-$0.12& $-$0.26&0.09&$-$0.11 &0.21&$-$0.79 \\
    \hline
  \end{tabular}
  \medskip 
  \end{center}
\end{table}

\vskip 0.7cm

Table  2

\end{center}

\newpage

\begin{center}
 
\begin{figure}[bth]
\centering
\centerline{
\epsfbox{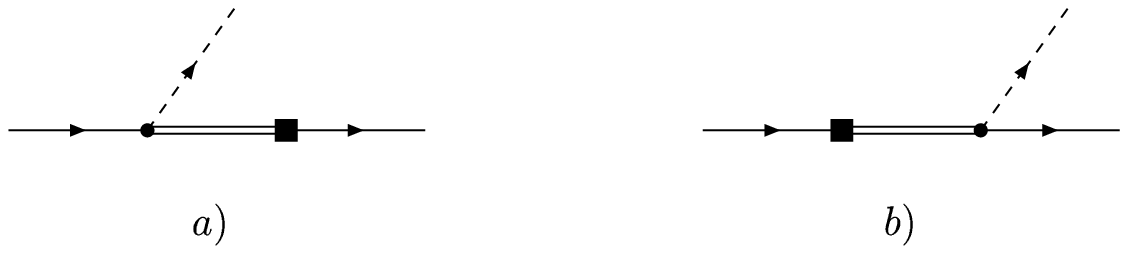}}
\end{figure}

\vskip 0.7cm

Figure 1

\vskip 1.5cm

\begin{figure}[tbh]
\centering
\centerline{
\epsfbox{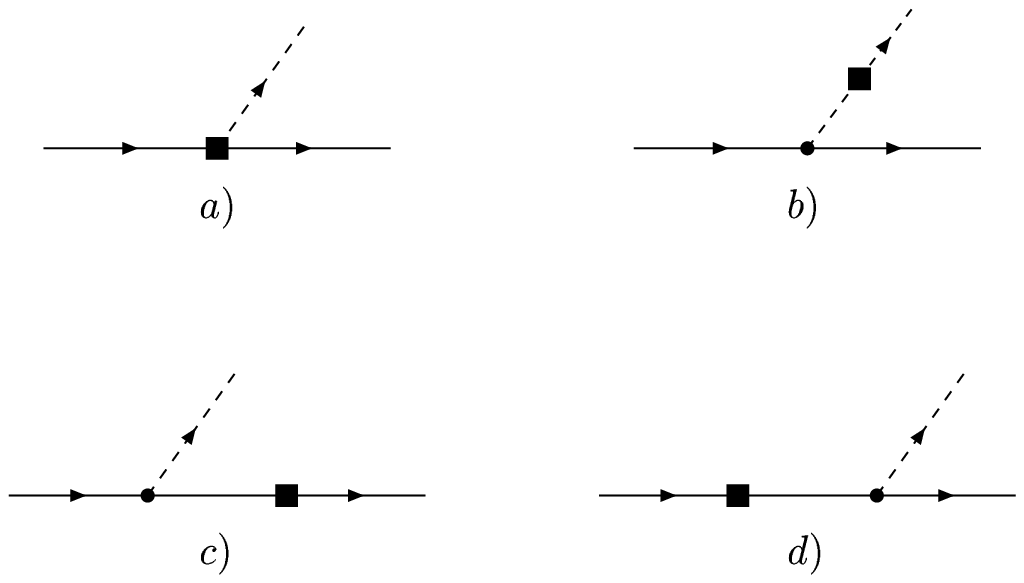}}
\end{figure}

\vskip 0.7cm

Figure 2

\end{center}

\end{document}